\documentclass{article}

\usepackage[nonatbib,preprint]{neurips_2019}

\usepackage[utf8]{inputenc} 
\usepackage[T1]{fontenc}    
\usepackage{hyperref}       
\usepackage{url}            
\usepackage{booktabs}       
\usepackage{amsfonts}       
\usepackage{nicefrac}       
\usepackage{microtype}      

\usepackage{tikz}
\usepackage{pgfplots}
\usepackage{amsmath}
\usepackage{amssymb}
\usepackage{wrapfig}
\usepackage{booktabs}
\usepackage{caption}
\usepackage{xcolor}
\usepackage{comment}
\usepackage{epsfig}
\usepackage{graphicx}
\usepackage{multirow}
\usepackage{tabularx}
\usepackage{xspace}
\newcolumntype{C}{>{\centering\arraybackslash}X}
\newcolumntype{R}{>{\raggedleft\arraybackslash}X}

\newcommand{\rivagan}{\textsc{RivaGAN}\xspace}

\newcommand{\conv}{\mathtt{Conv}}
\newcommand{\cat}{\mathtt{Cat}}
\newcommand{\pool}{\mathtt{Pool}}
\newcommand{\linear}{\mathtt{Linear}}
\renewcommand{\tanh}{\mathtt{TanH}}

\title{Robust Invisible Video Watermarking with Attention}
\author{
    Kevin Alex Zhang\\
    LIDS, MIT\\
    Cambridge, MA 02142\\
    \texttt{kevz@mit.edu}\\
   \And
    Lei Xu\\
    LIDS, MIT\\
    Cambridge, MA\\
    \texttt{leix@mit.edu}\\
    \AND
    Alfredo Cuesta-Infante\\
    Univ. Rey Juan Carlos\\
    Móstoles, Spain\\
    \texttt{alfredo.cuesta@urjc.es}\\
    \And
    Kalyan Veeramachaneni\\
    LIDS, MIT\\
    Cambridge, MA 02142\\
    \texttt{kalyanv@mit.edu}
}
\begin{document}

\maketitle

\begin{abstract}
The goal of video watermarking is to embed a message within a video file in a way such that it minimally impacts the viewing experience but can be recovered even if the video is redistributed and modified, allowing media producers to assert ownership over their content. This paper presents \rivagan, a novel architecture for robust video watermarking which features a custom attention-based mechanism for embedding arbitrary data as well as two independent adversarial networks which critique the video quality and optimize for robustness. Using this technique, we are able to achieve state-of-the-art results in deep learning-based video watermarking and produce watermarked videos which have minimal visual distortion and are robust against common video processing operations.
\end{abstract}


\section{Introduction}
Video watermarking is a set of techniques that aims to hide information in a video stream in such a way that it is hard to remove or tamper with, all while preserving the quality and fidelity of the content.  Video watermarking allows content creators to prove ownership after distribution and enables movie producers to identify leaks by embedding unique identifiers into preview copies of films \cite{OverviewOfDigitalWatermarking}. Other examples of applications include everything from identifying copyright infringement and embedding tags for filtering content to automated broadcast monitoring for commercials.


Effective video watermarking is both \textit{invisible} and \textit{robust}. However, existing techniques rarely achieve both of these goals at once. Invisible watermarking is much more challenging in videos than in still images because perturbing frames independently may result in highly visible distortions such as flickering. Also, classical watermarking techniques based on algorithms such as the discrete cosine transform or discrete wavelet transform are typically not robust to video processing operations like cropping and scaling. If the leaked video has undergone any of these geometric transformations, the watermark may be destroyed.  



The goal of this paper is to design a deep learning-based, multi-bit video watermarking process that is both robust and invisible. We are motivated by the recent success of deep learning and adversarial training methods in data hiding tasks as shown by \cite{HiDDeN,StegaStamp}. In this paper, we propose a novel architecture that goes beyond the standard convolutional layers and operations used in related deep learning based systems such as image steganography and watermarking.

Our paper is organized as follows: Section~\ref{sec:related_work} discusses related work in watermarking and steganography, Section~\ref{sec:model} introduces our approach to video watermarking, Section~\ref{sec:experiments} presents some results on benchmark datasets, and Section~\ref{sec:discussion} provides additional insights into how our model functions.

\section{Related Work}\label{sec:related_work}
Watermarking is closely related to steganography, as both aim to hide information within another medium. Recent work at the intersection of deep learning and steganography has shown promising results. For example, \cite{HiDDeN, Hayes} have explored using convolutional neural networks for image steganography and managed to achieve higher relative payloads. Additionally, \cite{ConvolutionalVideoSteganography} explores video steganography in the context of secretly embedding one video inside of another. While they both involve hiding data inside another piece of content, steganography assumes that adversaries are attempting to detect a message, while watermarking assumes that they are attempting to remove the watermark. 


Watermarking techniques are usually classified as image-based (operating on each frame independently) or video-based (exploiting temporal information). Image-based techniques operate either in the spatial or frequency domain \cite{ImageBasedReview}. Spatial domain methods include approaches such as modifying the least significant bit of each pixel which is a straightforward but weak method as the watermark could be easily removed by a random modification of all pixels. More elaborate proposals include Spread Spectrum Modulation \cite{SSM01, SSM02, SSM03} and Quantization Index Modulation \cite{QIM01, QIM02}. In the frequency domain, the watermark can be embedded by modifying the coefficients produced with any of the following transformations of the video sequence:
the Discrete Cosine Transform (DCT) \cite{DCT01},
the Discrete Fourier Transform \cite{DFT01},
and the Discrete Wavelet Transform \cite{DWT01, DWT02, DWT03}.

There are two main approaches for video-based techniques, the first of which is the compression format. For instance, Reversible Variable Length Codes (RVLC), introduced in \cite{CompressionFormat01}, can be based on the H.264 format \cite{CompressionFormat02} or on MPEG \cite{CompressionFormat03}. Alternatively, one might focus on the motion vectors produced by the compression process, slightly altering any with a magnitude large enough that the alteration will not be visible, as in \cite{MotionVector01, MotionVector02}.
Recently Convolutional Neural Networks (CNN) and Deep Learning (DL) have been used for image steganography \cite{ImageSteganDL01} -- suggesting that they may also be suitable for watermarking. For example, a secure signal authentication based on dynamic watermarking with DL was recently proposed in \cite{DeepWatermarking01}. Steganography with DL for still images has recently been investigated \cite{HiDDeN, Hayes} and used for hiding a video within another video in \cite{ConvolutionalVideoSteganography}. Other approaches to video watermarking such as perceptual models \cite{TexelWatermarking,BlockWatermarking}, object detection \cite{FragileObjectWatermarking,GeometricalObjectWatermarking}, and scene segmentation \cite{PerceptualModelsAndSceneSegmentation}, are also used in computer vision tasks where DL achieves state-of-the-art results. As far as we know, however, video watermarking with DL has only recently begun to be explored, and is still an open problem \cite{DeepWatermarking02, DeepWatermarking03}. In this paper, we propose an end-to-end model for robust video watermarking which uses multiple adversaries as well as a novel attention mechanism.



\section{RivaGAN}\label{sec:model}
In this section, we introduce our model for robust video watermarking. Our goal is to encode a $D$-bit data vector, where $D \in \{32, 64\}$, into an arbitrary video of $T$ frames in such a way that (1) the bit vector can be reliably recovered given one or more frames of the watermarked video,
(2) there are no visible distortions, (3) the watermark cannot be easily removed by watermark removal tools, and (4) the watermark is robust against common video processing operations. 

\begin{figure*}
\begin{center}
    \includegraphics[width=\linewidth]{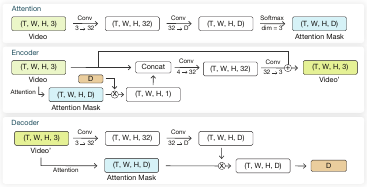}
\end{center}
\caption{This figure shows how the attention, encoder, and decoder modules operate on a tensor level. The attention module uses two convolutional blocks to create an attention mask, which is then used by the encoder and decoder modules to determine which bits to pay attention to at each pixel. The encoder module uses the attention mask to compute a compacted form of the data tensor and concatenates it to the image before applying additional convolutional blocks to generate the watermarked video. The decoder module extracts the data from each pixel but then weights the prediction using the attention mask before averaging to try and recover the original data.}
\label{fig:tensorflow}
\vspace{-0.5em}
\end{figure*}

To achieve these goals, we design our architecture with two adversaries: a \textit{critic} which evaluates the quality of the watermarked video, and an \textit{adversary} network which attempts to remove the watermark. These two work with an encoder network which adds the watermark to a video, and a decoder network which extracts the watermark.
We present our architecture in Figure~\ref{fig:tensorflow} and present details about various transformations in Section~\ref{sect:architecture}.

\begin{figure*}[t]
\centering
\includegraphics[width=0.9\linewidth]{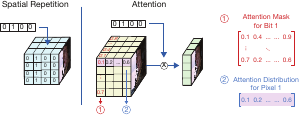}
\caption{This figure shows the difference between what related deep learning-based approaches (left) to this task use to represent their data and what our attention-based approach (right) uses. Unlike existing approaches which naively repeat the data cross the spatial dimensions, we learn a probability distribution over the data for each pixel (e.g. the attention distribution) and use that to generate a more compact data representation. This operation also has the advantage of being interpretable as an ``attention mask" as we can see what bits each pixel is paying attention to and encourage the model to pay attention to different bits based on the content of the image.}
\label{fig:attention}
\vspace{-0.5em}
\end{figure*}

In addition, we also introduce a new mechanism for combining the data and image representations which is more robust against common video processing operations and is easier to train. Currently, all existing approaches for deep learning-based data hiding techniques operate by concatenating the binary data to a feature map derived from the image and apply additional convolution layers to generate the output. We propose a different attention-based mechanism (shown in Figure~\ref{fig:attention} which learns a probability distribution over the data dimensions for each pixel and uses that distribution to select the bits to pay attention to during the embedding process. This biases our model towards learning to hide different bits in different objects and textures, making it easier to train and resulting in robustness against operations such as cropping, scaling, and compression.


\noindent \textbf{Notation.}
Let $X\in \mathbb{R}^{T\times W\times H\times C}$ be an tensor and $Y\in \mathbb{R}^{C'}$ be a vector. Then let $\cat: (X, Y)\rightarrow \Phi \in \mathbb{R}^{T\times W\times H\times (C+C')}$ be the concatenation of $X$ and $Y$, where $Y$ is expanded to a $T\times W\times H\times C'$ dimensional tensor.

Let $\conv_{D\rightarrow D'}: X\rightarrow \Phi$ be a 3D convolutional block that takes an input tensor $X \in \mathbb{R}^{T\times W\times H \times D}$ and maps it to a feature tensor $\Phi \in \mathbb{R}^{T\times W\times H \times D}$, where $T$, $W$ and $H$ are the time, width and height dimension respectively. $D$ and $D'$ are the depth of features. The convolutional block applies a $1 \times K \times K$ convolution kernel where $K=11$, followed by a $\tanh$ activation and a batch normalization operation \cite{BatchNormalization}.

Let $\pool: X\rightarrow \Phi$ be an adaptive mean pooling operation which takes an input tensor $X \in \mathbb{R}^{T\times W\times H \times D}$ and maps it to a feature tensor $\Phi \in \mathbb{R}^{D}$ by averaging over the $T$, $W$, and $H$ dimensions.

Let $\linear_{D\rightarrow D'}: X\in \mathbb{R}^{\ldots\times D} \rightarrow \Phi \in \mathbb{R}^{\ldots\times D'}$ be a linear transformation for the last dimension of a tensor. So $\Phi$ and $X$ are the same size except for the last dimension, which changes from $D$ to $D'$.

Finally, let $V$ and $\hat{V}$ be the original video and the watermarked video, which all have the same length and resolution $T\times W\times H$ and use RGB color space. Let $M\in\{0,1\}^{32}$ be the $32$-bit watermark, and $\hat{M}$ be the watermark recovered from $\hat{V}$. Let $\mathcal{T}$ be the attention module, let $\mathcal{E}$ be the encoder, let $\mathcal{D}$ be the decoder, let $\mathcal{C}$ be the critic, and let $\mathcal{A}$ be the adversary.

\subsection{Architecture}\label{sect:architecture}

\vspace{0.5em}\noindent\textbf{Attention.} The attention module is a pair of convolutional layers shared between the encoder and decoder. It takes the source frames, applies two convolutional blocks, and generates an attention mask of size $(T, W, H, D)$ where $D$ is the data dimension, and $(T, W, H)$ corresponds to the time and size dimensions. The attention mask functions by allowing the model to use the content of the image at a particular location to determine which dimensions of the data vector to pay attention to.


As shown in Figure~\ref{fig:tensorflow}, the output is an attention mask where the vector at each pixel can be interpreted as a multinomial distribution over the $D$ data dimensions. The attention module can be formally expressed as follows:
\begin{equation}
    \begin{cases}
        a &= \conv_{3\rightarrow 32}(V)\\
        b &= \conv_{32\rightarrow D}(a)\\
        \mathcal{T}(V) &= \mathtt{Softmax}(b)\\
    \end{cases}
\end{equation}

The attention mechanism biases the model towards hiding data in textures and objects that are less affected by these transformations than lower-level features. In this way, it helps encourage robustness against scaling and compression.
Empirically, we find that we are able to achieve faster convergence and better performance than with other approaches such as concatenation or multiplication as in \cite{HiDDeN}. We compare our attention-based approach to these competing approaches in Section~\ref{sec:experiments}.

\vspace{0.5em}\noindent\textbf{Encoder.} The encoder network is responsible for taking a fixed-length data vector and embedding it into a sequence of video frames. The encoder uses the attention module to generate a compact data tensor of shape $(T, W, H, 1)$, where the $D$ data dimensions have been reduced to a single real value using the attention weights, and concatenates this compact data tensor to the image. It then applies two convolutional blocks and generates a residual mask. We constrain the residual such that an individual pixel can be perturbed by no more than $\pm0.01$ and add the residual mask to the original video to generate the watermarked output\footnote{We represent pixel intensities as floating point numbers in the range $[-1.0, 1.0]$ as opposed to the discrete representation as integers in the set $\{0, ..., 255\}$.}. It can be formally expressed as follows: 
\begin{equation}
 \begin{cases}
        a &= \mathcal{T}(V) \times M\\
        b &= \conv_{4\rightarrow 32}(\cat(V, a))\\
        c &= \conv_{32\rightarrow 3}(b)\\
       \hat{V} &= \mathcal{E}(V, M) = V + 0.01 \cdot \tanh(c)
    \end{cases}
\end{equation}

Consider an extreme example: for a given pixel, the attention module generates an attention vector where all the values are 0 except in the first dimension, which is 1. In this case, the compacted data vector would simply contain the first bit of the data. This operation allows the encoder to learn to pay attention to different dimensions of the data conditioned on the content of the image at each pixel.


\vspace{0.5em}\noindent\textbf{Decoder.} The decoder network is responsible for taking a sequence of video frames and extracting the watermark. As shown in Figure~\ref{fig:tensorflow}, we (1) attempt to extract all $D$ bits of data from every location in the video and (2) reuse the attention module from the encoder, performing what we refer to as an ``attention pooling" operation to aggregate over the spatial dimensions and generate a $D$-dimensional prediction of the watermark bits. This operation can be formally expressed as:
\begin{equation}
    \begin{cases}
        a &= \conv_{3\rightarrow 32}(\hat{V})\\
        b &= \conv_{32\rightarrow D}(a)\\
        c &= \mathcal{T}(V) \times b\\
        \hat{M} &= \mathcal{D}(\hat{V}) = \pool(c)
    \end{cases}
\end{equation}

This operation is designed to take advantage of that fact that if the encoder paid a lot of attention to bit $d$ at a particular pixel, then the value of that pixel is more likely to contain information about bit $d$ than some arbitrary bit. Therefore we weigh the predictions generated by the decoder using the amount of attention paid to each bit at each location, and take the average.

We note that the decoder module does not require access to the original source video since the attention module is applied to the watermarked version of the video; as a result, this decoder satisfies the criteria for our system to be classified as a blind video watermarking algorithm.


\vspace{0.5em}\noindent\textbf{Critic.} The critic network is responsible for taking a sequence of video frames and detecting the presence of a watermark. It encourages the encoder to watermark the video in such a way that the distortion is less visible and can fool the critic. This module consists of two convolutional blocks, followed by an adaptive spatial pooling layer and a linear classification layer which produces the critic score.

\vspace{0.5em}\noindent\textbf{Adversary.} The adversary network attempts to imitate an attacker trying to remove the watermark. Specifically, the adversary network is responsible for taking a sequence of video frames and removing the watermark to generate another sequence of clean video frames. 
This module closely resembles the \textit{Encoder} module without the data tensor. This module consists of two convolutional blocks followed by a linear layer which generates the residual mask. We then apply a scaled $\tanh$ activation function to constrain the maximum amount by which an individual pixel can be perturbed to $\pm0.01$, and add the residual mask to the watermarked video to generate the output.

\subsection{Noise Layers}
In order to encourage robustness against common video transforms, we apply several \textit{noise} layers to the watermarked video before it is passed to the decoder, forcing the encoder and decoder to learn representations that are invariant to these transforms.

\vspace{0.5em}\noindent\textbf{Scaling.} The scaling layer is designed to re-scale the video to a random size where the width and height are between 80-100\% of the original. By inserting this noise layer between the encoder and decoder, we ensure that our model learns to embed data bits in a scale-invariant manner.

\vspace{0.5em}\noindent\textbf{Cropping.} The cropping layer is designed to randomly select a sub-window that contains 80-100\% of the video frame. By inserting this noise layer between the encoder and decoder, we ensure that our model learns to embed the data bits with sufficient spatial redundancy that cropping will not remove the message.

\vspace{0.5em}\noindent\textbf{Compression.} The compression layer uses the discrete cosine transform (DCT) to provide a differentiable approximation of video compression algorithms such as H.264 \cite{H264}. By converting the video into the YCrCb color space, applying the 3D DCT transform, zeroing out 0-10\% of the highest frequency components, applying the inverse DCT transform, and then converting the video back into the RGB color space, we can force our model to embed watermarks in a compression-resistant manner.


\subsection{Optimization}
\noindent\textbf{Loss Functions.} In order to train the encoder $\mathcal{E}$ and decoder $\mathcal{D}$ in our video watermarking model, we minimize the following loss functions. The cross-entropy loss between the bit vector and the decoded data
    \begin{equation*}
        \mathcal{L}_d = \mathbb{E}_{V,M} [\text{CrossEntropy}(M, \mathcal{D}(\mathcal{E}(V, M)))]
    \end{equation*}
The cross-entropy loss between the bit vector and the decoded data after the watermarked video is processed by the non-differentiable MJPEG compression operation. This operation takes the sequence of frames generated by the model, saves it to disk using the MJPEG compression format, and reads it back for the decoder to process. This loss can be expressed by
    \begin{equation*}
        \mathcal{L}_d^* = \mathbb{E}_{V,M} [\text{CrossEntropy}(M, \mathcal{D}(\text{MJPEG}(\mathcal{E}(V, M))))]
    \end{equation*}
The realism of the watermarked video according to the critic network
    \begin{equation*}
        \mathcal{L}_c = \mathbb{E}_{V,M} [\mathcal{C}(\mathcal{E}(V, M))]
    \end{equation*}
The cross-entropy loss between the bit vector and the data that is recovered from the watermarked video after the adversary has tampered with it
    \begin{equation*}
        \mathcal{L}_a = \mathbb{E}_{V, M} [\text{CrossEntropy}(M, \mathcal{D}(\mathcal{A}(\mathcal{E}(V, M))))]
    \end{equation*}

To optimize the critic $\mathcal{C}$ and adversary $\mathcal{A}$ modules, we also use the following loss functions. The Wasserstein loss to distinguish between source and watermarked videos
    \begin{equation*}
        \mathcal{L}_w = \mathbb{E}_{V}[ \mathcal{C}(V)] - \mathbb{E}_{V, M} [\mathcal{C}(\mathcal{E}(V, M))]
    \end{equation*}
The negative cross-entropy loss to teach the adversary to remove the watermark
\begin{equation*}
    \mathcal{L}_r = -\mathbb{E}_{V, M} [\text{CrossEntropy}(M, \mathcal{D}(\mathcal{A}(\mathcal{E}(V, M))))]
\end{equation*}

\noindent\textbf{Training Procedure.} We optimize these loss functions using the Adam optimizer with an initial learning rate of $10^{-3}$ which is decayed when the loss function plateaus; furthermore, we clip the critic weights to $[-0.1, 0.1]$ and train our model for $300$ epochs. During the training stage, we use standard data augmentation procedures including random horizontal flipping (where we flip all frames in a given video) and random cropping (where we select a random sub-image from all frames). We operate on batches of size $N=12$ and our procedure for generating the batches involves selecting $N/2$ videos from the training dataset and pairing each video with (1) a randomly generated bit vector $D$ and (2) the complement of that bit vector $\bar{D}$. We refer to these paired samples as \textit{Hamming vector pairs} to denote that the two bit vectors differ by a single bit. We find that this procedure results in faster convergence and improves model performance significantly.


\section{Experiments and Results}\label{sec:experiments}
\textbf{Setup.} To evaluate the effectiveness of our approach, we run a series of experiments on the Hollywood2 \cite{Hollywood2} data set which contains over 2500 short video clips extracted from movies and adds up to over 20 hours of video content. We generated a random 32-bit identifier and measured our model's ability to hide and recover the data in a variety of different scenarios. In order to evaluate the decoding accuracy, we quantize the pixel values to 8-bit integers and store them as a MJPEG video file. We also report the decoding accuracy after apply other video processing operations such as cropping to examine our robustness against these transforms.

\begin{table}[t]
\caption{This table shows the results for a model trained to embed 32 bits of data with or without our attention mechanism. We find that models trained with our attention masking and pooling operations outperform models trained without it and are significantly more robust against geometric transforms. The average PSNR of the attention-based models is 42.65 while the average PSNR of the models without attention is 42.73.}
\label{table:results_mini}
\vspace{0.5em}
\fontsize{10}{14}\selectfont
\begin{tabularx}{\linewidth}{rCCCCCC}
\textbf{Model}                    & \textbf{MJPEG} & \textbf{Cropped} & \textbf{Scaled} \\ \hline
No Attention                      & 0.595          & 0.588            & 0.589           \\
No Attention + Noise              & 0.973          & 0.970            & 0.915           \\
Attention                         & 0.997          & 0.981            & 0.985           \\
Attention + Noise                 & \textbf{0.997} & \textbf{0.995}   & \textbf{0.987}  \\
\end{tabularx}
\vspace{-1em}
\end{table}

\textbf{How do we compare to concatenation-based models?} As shown in Table~\ref{table:results_mini}, we find that our attention-based models outperform concatenation-based models such as \cite{HiDDeN}. In general, we find that attention-based models are more robust to compression, cropping, and scaling even when we do not explicitly use noise layers to encourage the model to be robust to these transformations. Furthermore, even when noise layers are used, we find that our attention-based models still outperform concatenation-based approaches.

\textbf{How effective is our approach?} We show some examples of video frames in Figure~\ref{fig:diffs} and note that the watermarked video does not contain any noticeable artifacts. Our results are presented in Table~\ref{table:results} which shows our image quality and our ability to recover the watermark for different model configurations and different video processing operations. 

We find that when the watermarked video is transmitted without modification, the receiver is able to decode the 32-bit watermark with above 95\% accuracy in all cases. We note that this low error rate can easily be compensated for through error correcting codes, allowing our system to be used in real-world applications. Furthermore, we find that the cropping and scaling noise layers are effective at encouraging robustness against the corresponding video processing operations. When these layers are applied, the receiver is able to decode the watermark with approximately 99\% accuracy despite cropping and scaling.

\begin{table}[t]
\caption{This table shows the video quality and watermarking accuracy when embedding $D$ bits of random data into videos from the test set. The \textit{MJPEG} column indicate the accuracy obtained by the decoder after the video is compressed, saved, and read back. The \textit{Cropped} column indicates the accuracy obtained after the video is randomly cropped down to 80\% of its original size, compressed, saved, and read back. Similarly, the \textit{Scaled} column indicates the accuracy obtained after the video is randomly scaled down to 80\% of its original size, compressed, saved, and read back.}
\label{table:results}
\vspace{0.5em}
\fontsize{10}{14}\selectfont
\begin{tabularx}{\linewidth}{rCCCCCC}
\multicolumn{1}{c}{}              & \multicolumn{1}{l}{} & \multicolumn{2}{c}{\textbf{Quality}} & \multicolumn{3}{c}{\textbf{Accuracy}}                \\
\textbf{Model}                    & \textbf{D}            & \textbf{PSNR}     & \textbf{SSIM}    & \textbf{MJPEG} & \textbf{Cropped} & \textbf{Scaled} \\ \hline
Attention                              & 32                    & 42.71             & 0.954            & 0.997          & 0.981            & 0.985           \\
Attention + Noise                      & 32                    & 42.61             & 0.960            & 0.997          & 0.995            & 0.987           \\
Attention + Noise + Critic             & 32                    & 42.08             & 0.948            & 0.998          & 0.998            & 0.991           \\
Attention + Noise + Critic + Adversary & 32                    & 42.05             & 0.960            & 0.992          & 0.988            & 0.981           \\ \hline
Attention                              & 64                    & 42.20             & 0.944            & 0.993          & 0.980            & 0.961           \\
Attention + Noise                      & 64                    & 42.22             & 0.953            & 0.971          & 0.966            & 0.917           \\
Attention + Noise + Critic             & 64                    & 42.06             & 0.945            & 0.991          & 0.989            & 0.961           \\
Attention + Noise + Critic + Adversary & 64                    & 41.99             & 0.950            & 0.983          & 0.972            & 0.958
\end{tabularx}
\end{table}

\textbf{Can humans identify the watermarked video?} To further establish the invisibility of our watermarking scheme, we asked workers on the Mechanical Turk platform to watch a random selection of videos and try to distinguish the source videos from the watermarked videos. For this experiment, we generated pairs of source and watermarked videos for all 884 videos in our test set and asked workers who possessed the ``masters" qualification to review each pair and identify which video contained the watermark.

\begin{wraptable}{r}{0.6\textwidth}
    \vspace{-1em}
    \caption{This table shows the detection rate by workers on Mechanical Turk for a randomly selected subset of test videos generated by each model.}
    \label{table:mturk}
    \vspace{0.5em}
    \centering
    \begin{tabularx}{\linewidth}{Rc}
        \toprule
        Model & Detection Rate \\
        \midrule
        Attention + Noise & 0.541 \\
        Attention + Noise + Critic & 0.514 \\
        Attention + Noise + Critic + Adversary & 0.515 \\
        \bottomrule
    \end{tabularx}
    \vspace{-1em}
\end{wraptable}

We present the results of this experiment in Table~\ref{table:mturk} and note that the human workers are only slightly better than random guessing. Furthermore, we find evidence to suggest that the critic module reduces the visibility of the watermark as the detection rate for watermarked videos generated by the critic model is 5\% lower than those generated by the baseline models.


\section{Additional Insights}\label{sec:discussion}

\begin{figure*}[t!]
\centering
\includegraphics[width=0.195\linewidth]{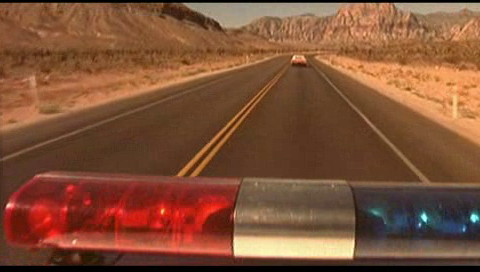}
\includegraphics[width=0.195\linewidth]{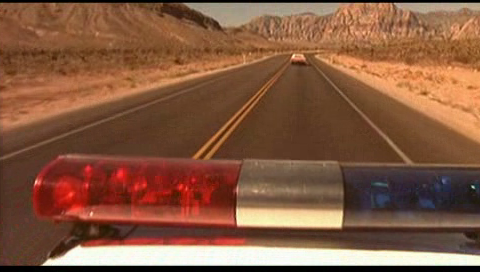}
\includegraphics[width=0.195\linewidth]{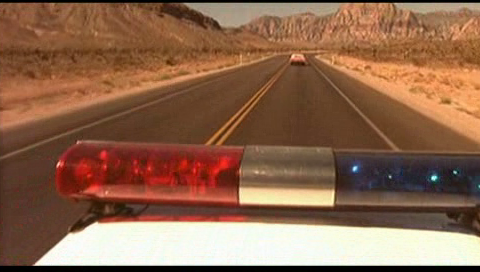}
\includegraphics[width=0.195\linewidth]{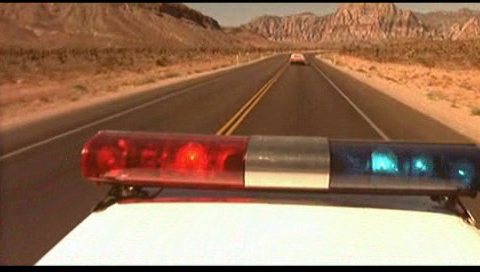}
\includegraphics[width=0.195\linewidth]{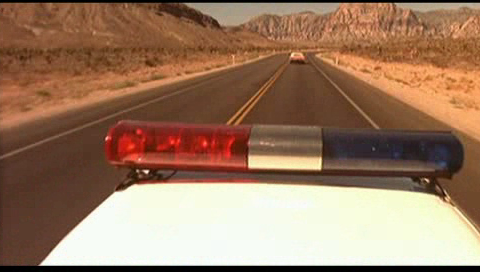}
\hfil
\includegraphics[width=0.195\linewidth]{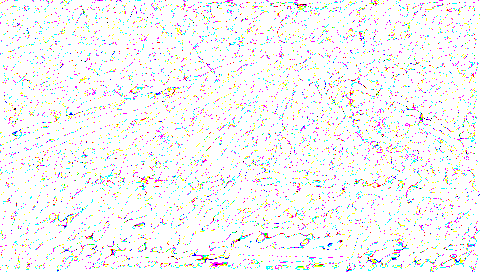}
\includegraphics[width=0.195\linewidth]{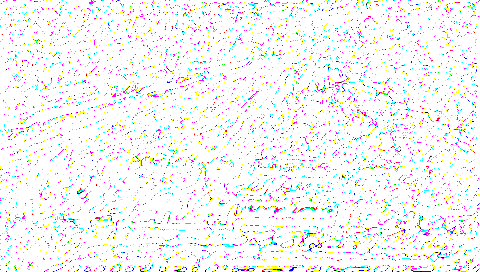}
\includegraphics[width=0.195\linewidth]{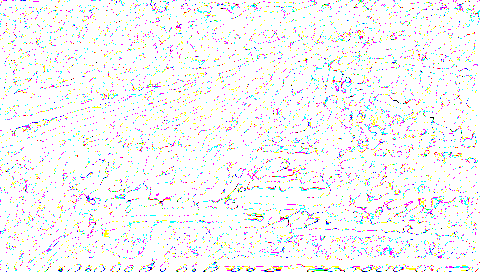}
\includegraphics[width=0.195\linewidth]{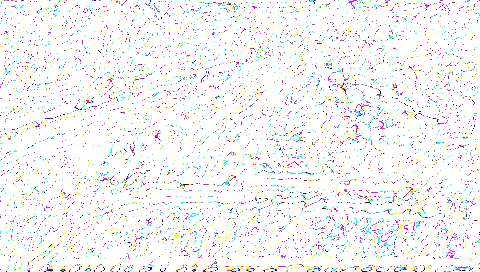}
\includegraphics[width=0.195\linewidth]{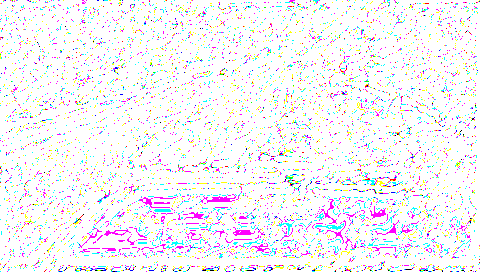}
\caption{This figure shows the watermarked video (top) and the residual masks (bottom). The residual masks were generated by the encoder module and added to the source video to produce the watermarked video.}
\label{fig:diffs}
\vspace{-0.5em}
\end{figure*}

\textbf{What does the watermark look like?} Next, we'll examine where the watermark data is being hidden by visually inspecting the \textit{residual} that is generated by the encoder and added to the source video. Figure \ref{fig:diffs} shows an example of a source video and the corresponding residuals. We note that the residual values appear to be fairly evenly distributed across the frame.

\textbf{How does changing a single bit change the watermark?} Finally, we examine the impact of flipping a single bit in the data tensor by examining the resulting ``distance mask". We compute each distance mask by taking a fixed data tensor $D_1$, embedding it in the image to generate a watermarked video $W_1$, changing a single bit in $D_1$ to create $D_2$, and embedding it in the image to generate a watermarked video $W_2$. Then, we visualize the difference between the two watermarked videos $|W_1-W_2|$ to highlight the regions of the watermarked video are affected by that particular bit.

\begin{figure}[t!]
\centering
\includegraphics[width=0.328\linewidth]{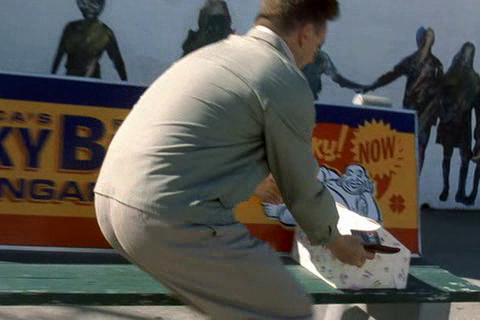}
\includegraphics[width=0.328\linewidth]{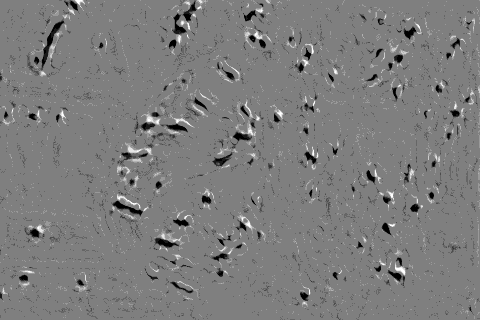}
\includegraphics[width=0.328\linewidth]{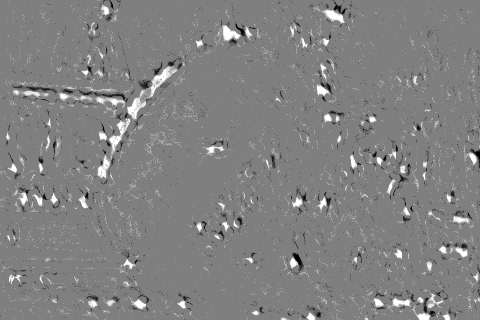}
\includegraphics[trim={0 0 0 42px},clip,width=0.328\linewidth]{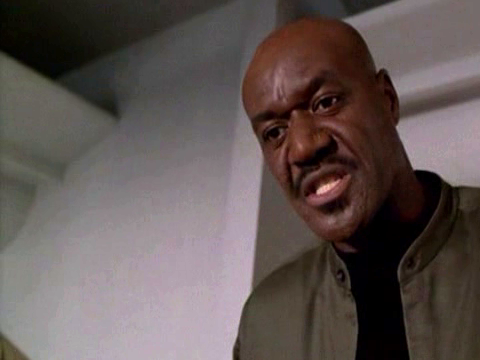}
\includegraphics[trim={0 0 0 42px},clip,width=0.328\linewidth]{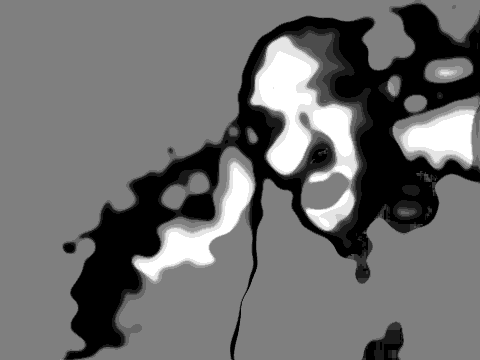}
\includegraphics[trim={0 0 0 42px},clip,width=0.328\linewidth]{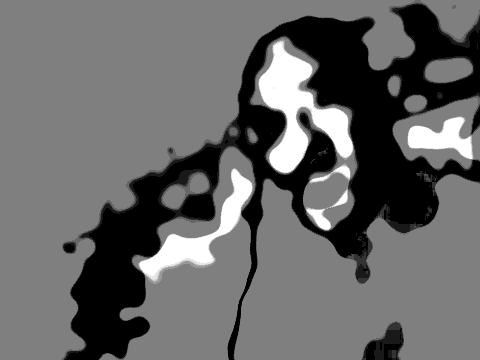}
\caption{This figure shows the original source video and two examples ``difference masks" for the first and second bit of the data tensor. Bright regions indicate that flipping a single bit caused that pixel to change in the watermarked output. The three images on the top correspond to a model trained with the attention mechanism and we note that the two difference masks look significantly different. The three images on the bottom correspond to a model trained without the attention mechanism and the two difference masks are virtually identical.}
\label{fig:bit_flip} 
\vspace{-0.5em}
\end{figure}

We perform this process with a randomly selected image for the first and second bits in our data tensor and present the results in Figure~\ref{fig:bit_flip}. This figure provides evidence to support that our hypothesis that the attention mechanism allows our model to pay attention to different dimensions of the data tensor depending on the content of the image as different bits appear to affect different parts of the watermark. We observe that the difference masks for the two bits are significantly different in the attention-based model but are not significantly different in the model without attention, suggesting that this phenomena can be attributed to the attention mechanism.

\begin{figure}[!tb]
\begin{center}
\begin{tikzpicture}[scale=0.6]
\begin{axis}[xlabel={Wall Clock Time},ylabel={Training Loss}]
\addplot[only marks,color=blue,mark=*] table[x expr=\thisrow{epoch}*2, y=train.loss, col sep=tab] {examples/bit_inverse/bit_inverse_big_lr.tsv};
\addplot[only marks,color=red,mark=*] table[x expr=\thisrow{epoch}*2, y=train.loss, col sep=tab] {examples/bit_inverse/bit_inverse_small_lr.tsv};
\addplot[only marks,color=green,mark=*] table[x=epoch, y=train.loss, col sep=tab] {examples/bit_inverse/no_inverse_big_lr.tsv};
\addplot[only marks,color=orange,mark=*] table[x=epoch, y=train.loss, col sep=tab] {examples/bit_inverse/no_inverse_small_lr.tsv};
\end{axis}
\end{tikzpicture}
\quad
\begin{tikzpicture}[scale=0.6]
\begin{axis}[xlabel={Wall Clock Time},ylabel={Test Accuracy},legend pos=outer north east,legend cell align={left}]
\addplot[only marks,color=blue,mark=*] table[x expr=\thisrow{epoch}*2, y=val.mjpeg_acc, col sep=tab] {examples/bit_inverse/bit_inverse_big_lr.tsv};
\addplot[only marks,color=red,mark=*] table[x expr=\thisrow{epoch}*2, y=val.mjpeg_acc, col sep=tab] {examples/bit_inverse/bit_inverse_small_lr.tsv};
\addplot[only marks,color=green,mark=*] table[x=epoch, y=val.mjpeg_acc, col sep=tab] {examples/bit_inverse/no_inverse_big_lr.tsv};
\addplot[only marks,color=orange,mark=*] table[x=epoch, y=val.mjpeg_acc, col sep=tab] {examples/bit_inverse/no_inverse_small_lr.tsv};
\legend{{Hamming Vector (LR = 0.001)}, {Hamming Vector (LR = 0.0001)}, {No Hamming Vector (LR = 0.001)}, {No Hamming Vector (LR = 0.0001)}}
\end{axis}
\end{tikzpicture}
\caption{This figure shows the training loss for the same model architecture, learning rate, and optimizer but trained with and without the bit inverse trick. We find that including the bit inverse within the same batch results in dramatically faster convergence as well as better model performance.}
\label{fig:bit_inverse}
\end{center}
\end{figure}
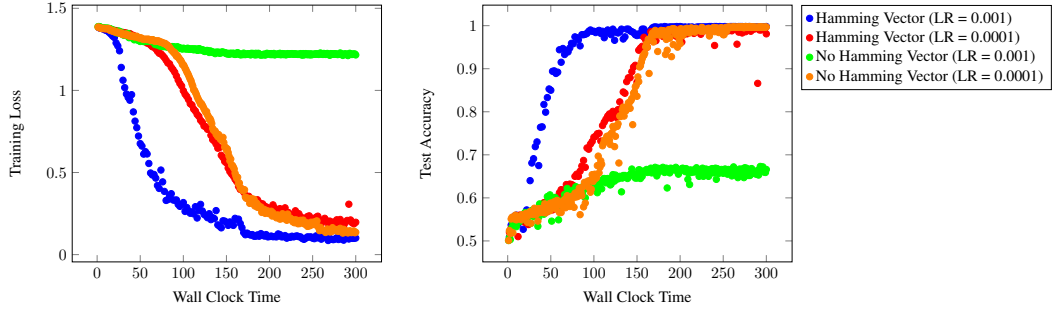

\textbf{Do we need to use Hamming vector pairs?} In our initial explorations, we trained our model by iterating over all of the videos in our dataset and training our model to encode and decode a randomly generated bit vector into each video. Despite experimenting with multiple optimizers, batch sizes, and learning rates, we found that our model often failed to converge within a reasonable number of epochs. This is shown in Figure~\ref{fig:bit_inverse} where the model trained with a high learning rate and without the Hamming vector pairs fail to converge.

In order to overcome this instability, we introduced the concept of Hamming vector pairs and found that the model converges significantly faster and, in the case of a high initial learning rate, achieves higher test accuracy. We hypothesize that this is due to the fact that the gradients produced by Hamming vector pairs are less noisy than the gradients produced by a simple random sample.

\section{Conclusion}
In this paper, we introduced a new class of attention-based architectures for data hiding tasks such as steganography and watermarking which is superior to existing approaches such as \cite{HiDDeN} as it (1) uses less memory, (2) is easier to train, and (3) is robust against common video processing operations such as scaling, cropping, and compression. We demonstrated the effectiveness of our approach on the video watermarking task, achieving near perfect accuracy with minimal visual distortion when hiding an arbitrary 32-bit watermark into video files. Our code is publically available and can be found online at: \url{https://github.com/DAI-Lab/RivaGAN}.

\bibliographystyle{plain}
\bibliography{egbib}

\begin{thebibliography}{10}

\bibitem{SSM02}
H.~O. {Altun}, A.~{Orsdemir}, G.~{Sharma}, and M.~F. {Bocko}.
\newblock Optimal spread spectrum watermark embedding via a multistep
  feasibility formulation.
\newblock {\em IEEE Trans. on Image Processing}, 18(2):371--387, Feb 2009.

\bibitem{OverviewOfDigitalWatermarking}
M.~Asikuzzaman and M.~R. Pickering.
\newblock An overview of digital video watermarking.
\newblock {\em IEEE Transactions on Circuits and Systems for Video Technology},
  28(9):2131--2153, Sep. 2018.

\bibitem{BlockWatermarking}
Zhila Bahrami and Fardin~Akhlaghian Tab.
\newblock A new robust video watermarking algorithm based on surf features and
  block classification.
\newblock {\em Multimedia Tools and Applications}, 77(1):327--345, 2018.

\bibitem{ImageSteganDL01}
Shumeet Baluja.
\newblock {Hiding Images in Plain Sight: Deep Steganography}.
\newblock In {\em {Proc. of the Conf. on Neural Information Processing Systems
  (NIPS) }}, 2017.

\bibitem{CompressionFormat03}
S.~{Biswas}, S.~R. {Das}, and E.~M. {Petriu}.
\newblock An adaptive compressed mpeg-2 video watermarking scheme.
\newblock {\em {IEEE Trans. on Instrumentation and Measurement}},
  54(5):1853--1861, Oct 2005.

\bibitem{DeepWatermarking01}
A.~{Ferdowsi} and W.~{Saad}.
\newblock Deep learning-based dynamic watermarking for secure signal
  authentication in the internet of things.
\newblock In {\em {Proc. of the IEEE Int. Conf. on Communications (ICC)}},
  pages 1--6, May 2018.

\bibitem{ImageBasedReview}
Garima Gupta, V.~K. Gupta, and Mahesh Chandra.
\newblock Review on video watermarking techniques in spatial and transform
  domain.
\newblock In Suresh~Chandra Satapathy, Jyotsna~Kumar Mandal, Siba~K. Udgata,
  and Vikrant Bhateja, editors, {\em Information Systems Design and Intelligent
  Applications}, pages 683--691, 2016.

\bibitem{SSM01}
Frank Hartung and Bernd Girod.
\newblock Watermarking of uncompressed and compressed video.
\newblock {\em Signal Processing}, 66(3):283 -- 301, 1998.

\bibitem{Hayes}
Jamie Hayes and George Danezis.
\newblock Generating steganographic images via adversarial training.
\newblock In {\em NIPS}, 2017.

\bibitem{FragileObjectWatermarking}
Dajun He, Qibin Sun, and Qi~Tian.
\newblock A semi-fragile object based video authentication system.
\newblock In {\em Proc. of the 2003 Int. Symposium on Circuits and Systems},
  volume~3, 2003.

\bibitem{DCT01}
J.~R. {Hernandez}, M.~{Amado}, and F.~{Perez-Gonzalez}.
\newblock Dct-domain watermarking techniques for still images: detector
  performance analysis and a new structure.
\newblock {\em IEEE Transactions on Image Processing}, 9(1):55--68, Jan 2000.

\bibitem{BatchNormalization}
Sergey {Ioffe} and Christian {Szegedy}.
\newblock {Batch Normalization: Accelerating Deep Network Training by Reducing
  Internal Covariate Shift}.
\newblock {\em arXiv e-prints}, page arXiv:1502.03167, Feb 2015.

\bibitem{QIM01}
Nie Jie and Wei Zhiqiang.
\newblock A new public watermarking algorithm for rgb color image based on
  quantization index modulation.
\newblock In {\em 2009 Int. Conf. on Information and Automation}, pages
  837--841, June 2009.

\bibitem{DWT03}
S.~{Kadu}, C.~{Naveen}, V.~R. {Satpute}, and A.~G. {Keskar}.
\newblock Discrete wavelet transform based video watermarking technique.
\newblock In {\em {Proc. of the Int. Conf. on Microelectronics, Computing and
  Communications (MicroCom)}}, pages 1--6, Jan 2016.

\bibitem{QIM02}
N.~K. {Kalantari} and S.~M. {Ahadi}.
\newblock A logarithmic quantization index modulation for perceptually better
  data hiding.
\newblock {\em IEEE Transactions on Image Processing}, 19(6):1504--1517, June
  2010.

\bibitem{DeepWatermarking03}
Haribabu Kandi, Deepak Mishra, and Subrahmanyam R.K.~Sai Gorthi.
\newblock Exploring the learning capabilities of convolutional neural networks
  for robust image watermarking.
\newblock {\em Computers \& Security}, 65:247 -- 268, 2017.

\bibitem{DWT02}
Ashish~M. Kothari and Ved~Vyas Dwivedi.
\newblock Transform domain video watermarking: Design, implementation and
  performance analysis.
\newblock In {\em {Proc. of the Int. Conf. on Communication Systems and Network
  Technologies}}, pages 133--137, 2012.

\bibitem{GeometricalObjectWatermarking}
Jung-Soo Lee and Whoi-Yul Kim.
\newblock A new object-based image watermarking robust to geometrical attacks.
\newblock In {\em Pacific-Rim Conference on Multimedia}, pages 58--64.
  Springer, 2004.

\bibitem{SSM03}
S.~P. {Maity} and S.~{Maity}.
\newblock Multistage spread spectrum watermark detection technique using fuzzy
  logic.
\newblock {\em IEEE Signal Processing Letters}, 16(4):245--248, April 2009.

\bibitem{Hollywood2}
Marcin Marsza{\l}ek, Ivan Laptev, and Cordelia Schmid.
\newblock Actions in context.
\newblock In {\em IEEE Conference on Computer Vision \& Pattern Recognition},
  2009.

\bibitem{CompressionFormat01}
Bijan~G. Mobasseri and Domenick Cinalli.
\newblock Reversible watermarking using two-way decodable codes.
\newblock In {\em {Proc. of the Int. Society for Optical Engineering, Security,
  Steganography, and Watermarking of Multimedia (VI)}}, pages 397--404, 2004.

\bibitem{MotionVector02}
N.~{Mohaghegh} and O.~{Fatemi}.
\newblock H.264 copyright protection with motion vector watermarking.
\newblock In {\em {Int. Conf. on Audio, Language and Image Processing}}, pages
  1384--1389, July 2008.

\bibitem{CompressionFormat02}
M.~{Noorkami} and R.~M. {Mersereau}.
\newblock A framework for robust watermarking of h.264-encoded video with
  controllable detection performance.
\newblock {\em IEEE Trans. on Information Forensics and Security}, 2(1):14--23,
  March 2007.

\bibitem{DFT01}
S.~{Pereira}, J.~J. K.~O. {Ruanaidh}, F.~{Deguillaume}, G.~{Csurka}, and
  T.~{Pun}.
\newblock Template based recovery of fourier-based watermarks using log-polar
  and log-log maps.
\newblock In {\em {Proc. of the IEEE Int. Conf. on Multimedia Computing and
  Systems}}, volume~1, pages 870--874, June 1999.

\bibitem{H264}
Iain~E. Richardson.
\newblock {\em The H.264 Advanced Video Compression Standard}.
\newblock Wiley Publishing, 2nd edition, 2010.

\bibitem{TexelWatermarking}
Mathias Schlauweg, Dima Pr\"{o}frock, Benedikt Zeibich, and Erika M\"{u}ller.
\newblock Self-synchronizing robust texel watermarking in gaussian scale-space.
\newblock In {\em Proceedings of the 10th ACM Workshop on Multimedia and
  Security}, MM\&\#38;Sec '08, pages 53--62, New York, NY, USA, 2008. ACM.

\bibitem{PerceptualModelsAndSceneSegmentation}
M.~D. Swanson, Bin Zhu, B.~Chau, and A.~H. Tewfik.
\newblock Multiresolution video watermarking using perceptual models and scene
  segmentation.
\newblock In {\em Proceedings of International Conference on Image Processing},
  volume~2, pages 558--561 vol.2, Oct 1997.

\bibitem{StegaStamp}
Matthew Tancik, Ben Mildenhall, and Ren Ng.
\newblock Stegastamp: Invisible hyperlinks in physical photographs.
\newblock {\em CoRR}, abs/1904.05343, 2019.

\bibitem{DeepWatermarking02}
V.~{Vukoti\'{c}}, V.~{Chappelier}, and T.~{Furon}.
\newblock {Are Deep Neural Networks good for blind image watermarking?}
\newblock In {\em {Proc. of the IEEE Int. Workshop on Information Forensics and
  Security (WIFS)}}, pages 1--7, Dec 2018.

\bibitem{ConvolutionalVideoSteganography}
Xinyu Weng, Yongzhi Li, Lu~Chi, and Yadong Mu.
\newblock Convolutional video steganography with temporal residual modeling.
\newblock {\em CoRR}, abs/1806.02941, 2018.

\bibitem{MotionVector01}
B.~{Yann}, L.~{Nathalie}, and D.~{Jean-Luc}.
\newblock A comparative study of different modes of perturbation for video
  watermarking based on motion vectors.
\newblock In {\em {Proc. of the 12th Euro. Signal Processing Conf.}}, pages
  1501--1504, 2004.

\bibitem{HiDDeN}
Jiren Zhu, Russell Kaplan, Justin Johnson, and Li~Fei{-}Fei.
\newblock {HiDDeN: Hiding Data With Deep Networks}.
\newblock In {\em Proc. 15th Euro. Conf. on Computer Vision {(ECCV)} Part
  {XV}}, pages 682--697, 2018.

\bibitem{DWT01}
Wenwu Zhu, Zixiang Xiong, and Ya-Qin Zhang.
\newblock Multiresolution watermarking for images and video.
\newblock {\em IEEE Trans. on Circuits and Systems for Video Technology},
  9(4):545--550, June 1999.

\end{thebibliography}

\end{document}